\documentclass[11pt,twoside]{article}
\usepackage{asp2010}
\resetcounters

\begin{document}

\title{Is there plenty of metal-poor stars with planets in the Galactic thick disk?}
\author{V.~Zh.~Adibekyan,$^1$ N.~C.~Santos,$^{1,2}$ S.~G.~Sousa$^{1,3}$ G.~Israelian$^{3,4}$ and P.~Figueira$^{1}$
\affil{$^1$Centro de Astrof\'{\i}�sica da Universidade do Porto, Rua das Estrelas,
4150-762 Porto, Portugal -- \textsl{Vardan.Adibekyan@astro.up.pt}} 
\affil{$^2$Departamento de F\'{\i}�sica
e Astronomia, Faculdade de Ci\^{e}ncias da Universidade do Porto, Portugal}
\affil{$^3$Instituto de Astrof\'{\i}sica de Canarias, 38200 La Laguna, Tenerife, Spain}
\affil{$^4$Departamento de Astrof{\'\i}sica, Universidad de La Laguna, 38206 La Laguna, Tenerife, Spain}}

\begin{abstract}

We performed an uniform spectroscopic analysis of 1111 FGK dwarfs
observed as part of the HARPS GTO planet search program. We applied a purely
chemical approach, based on [$\alpha$/Fe] ratio, to distinguish the various stellar components
in the Galaxy. Apart from the well known Galactic thick and thin disks, we separated
an $\alpha$-enhanced stellar family at super-solar metallicities. The metal-rich high-$\alpha$ stars
have orbits similar to the thin disk stars, but they are similar to thick disk stars in terms
of age. Our data indicate that the incidence of stars with planets are greater among the
chemically separated “thick” disk stars with [Fe/H] $\lesssim$ -0.3 dex than they are among
“thin” disk stars in the same [Fe/H] interval. Our results allow us to suppose that a certain
chemical composition, and not the Galactic birth place of the stars, is the causative
factor for that.
\end{abstract}

\section{Introduction}
Shortly, after the discovery of the first extra-solar planet \citet{Gonzalez_98}, based on a small sample from 8 
planet-host stars (PHS), suggested that PHSs tend to be metal-rich compared with the nearby field FGK stars 
which are known to hold non-planet. The metal rich nature of the PHSs was confirmed in subsequent papers 
\citep[e.g.][] {Santos_01,Santos_04}. Although theoretical modeling suggests that metallicity is a key parameter 
of planet formation, \citet{Haywood_08,Haywood_09} studying the memberships of PHSs to different stellar 
populations, proposed that the presence of giant planets might be primarily a function of a parameter 
linked to galactocentric radius (density of molecular hydrogen), but not metallicity and the apparent correlation 
between metallicity and the detection of planets is a natural consequence of that. 
If the rate of giant planets does not depend on metallicity, then the core accretion theory of planet formation 
\citep[e.g.][] {Ida_04} loses its most important observational support, and it would lend support for the gravitational 
instability theory \citep[e.g.][] {Boss_01}. 

In this proceeding we present the results for a uniform spectroscopic analysis of 1111 FGK dwarfs in order to study 
the frequency of the PHSs in different stellar populations.

\section{The sample and elemental abundances}

The sample used in this work consists of 1111 FGK stars observed within the context of the HARPS GTO programs. 
The spectra have a resolution of R $\sim$ 110000 and signal-to-noise (S/N) ratio ranging from ~20 to ~2000. 
Fifty five percent of the spectra have S/N higher than 200 and about 16\% of stars have S/N lower than 100.

Precise stellar parameters for all the stars were taken from \citet[and references therein] {Sousa_11}. 
Elemental abundances of refractory elements are determined for the sample stars using a differential LTE analysis relative to 
the Sun with the 2010 revised version of the spectral synthesis code MOOG and a grid of Kurucz ATLAS9 atmospheres 
\citep[see][for details]{Adibekyan_12}.

\section{A new $\alpha$-enhanced super-solar metallicity population}

Left panel of Fig.~\ref{figure_1} shows [$\alpha$/Fe] versus [Fe/H] for the stars with \emph{$T{}_{\mathrm{eff}}$} =\emph{$T{}_{\mathrm{\odot}}$}
$\pm$ 500 K (''$\alpha$`` refers 
to the average abundance of Mg, Si and Ti). As can be seen, the stars are clearly separated into two groups 
according to the content of $\alpha$ elements: the ''high-$\alpha$`` and the ''low-$\alpha$'' stars (``thin'' disk). 
This separation highlights the well-known $\alpha$ enhancement of thick disk stars relative to the thin disk found for 
stars with [Fe/H] $<$ 0  \citep[e.g.][]{Bensby_03}. It is interesting to see that high-$\alpha$ stars are also divided into two 
subgroups: high-$\alpha$ metal-rich stars (hereafter ``h$\alpha$mr``), and high-$\alpha$ metal-poor stars (``thick'' disk).
The kinematic separation criteria suggest \citep[see][for details]{Adibekyan_11} that most of the stars in the h$\alpha$mr 
stellar family, such as chemically defined ''thin`` disk stars, have thin disk kinematics (see right panel of Fig.~\ref{figure_1}).

\articlefiguretwo{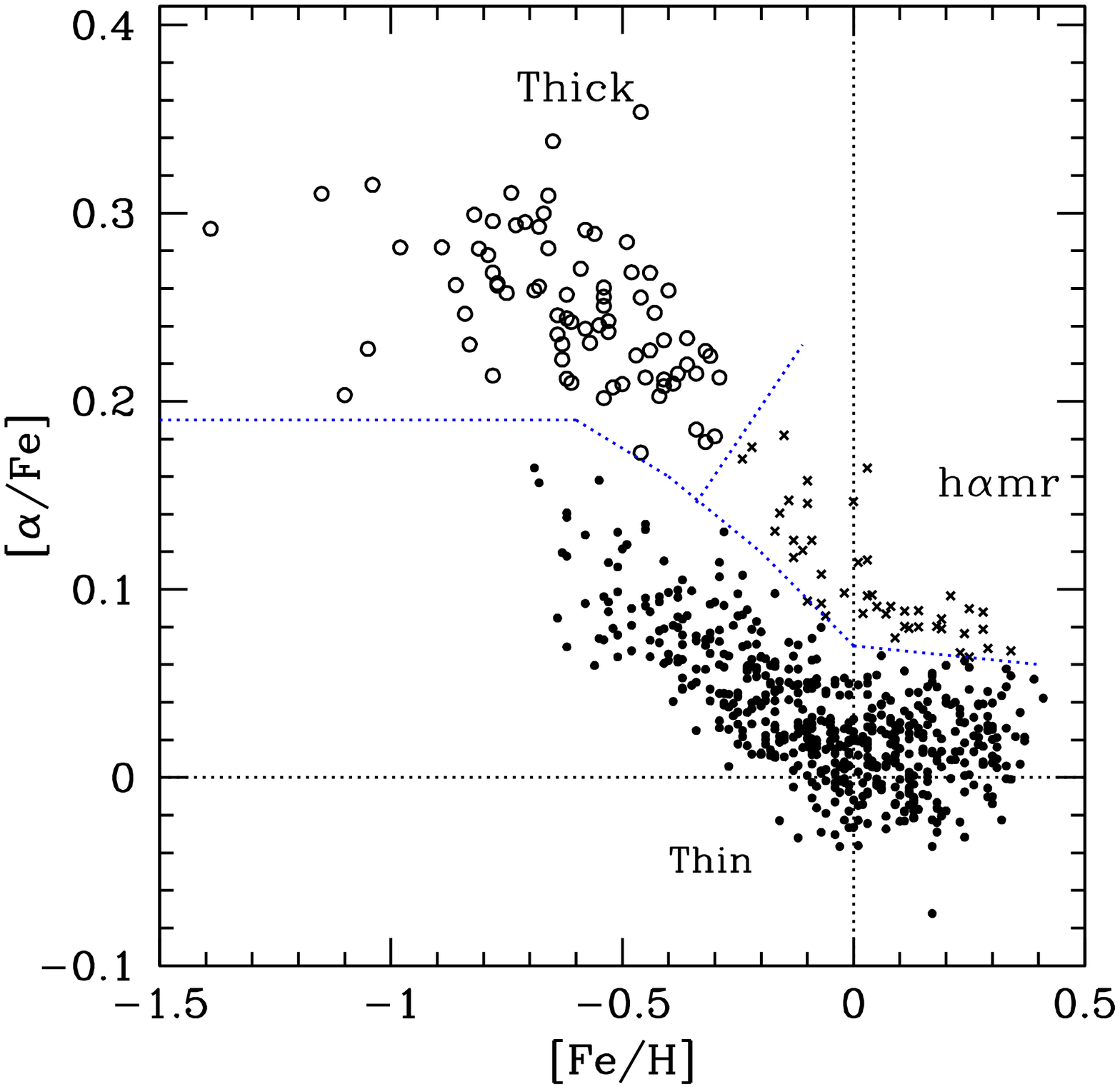}{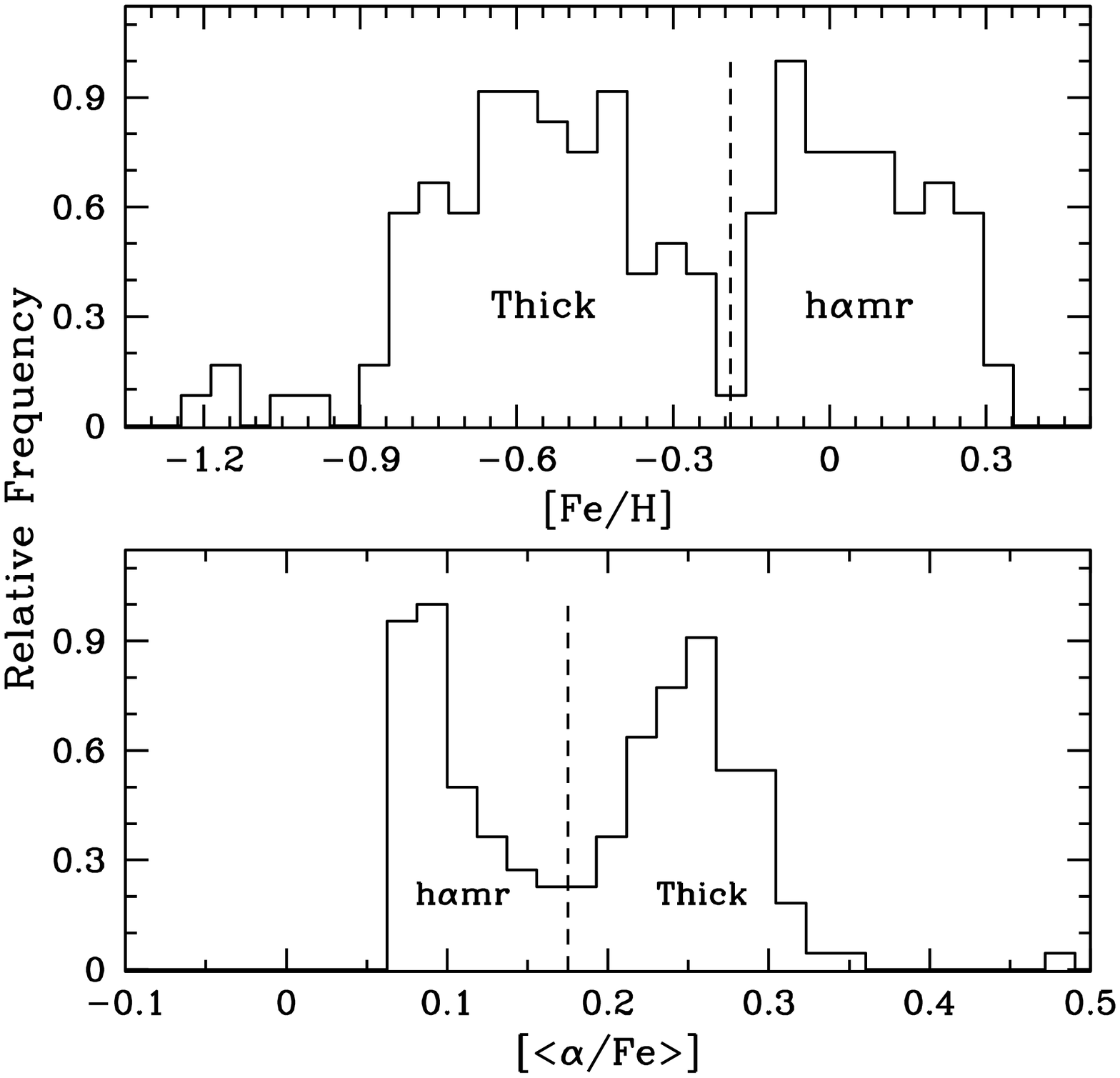}{figure_1}{\textit{Left} -- [$\alpha$/Fe] versus [Fe/H] for all stars with 
\emph{$T{}_{\mathrm{eff}}$} =\emph{$T{}_{\mathrm{\odot}}$} $\pm$ 500 K. 
Open circles refer to ''thick'' disk and dotes refer to ``thin'' disk stars, respectively. Crosses refer to the metal-rich high-$\alpha$ stars.
\textit{Right} --The [Fe/H] and [$\alpha$/Fe] separation histograms for the $\alpha$-enhanced stars.}


Studying the orbital parameters and ages of the three chemically separated groups we observe that metal-rich 
high-$\alpha$ (h$\alpha$mr) and the ''thick`` disk stars are on average older than chemically defined ''thin`` disk stars. 
Simultaneously the h$\alpha$mr stars, such as ''thin`` disk stars have nearly circular orbits, close to the Galactic plane.
Although the present observations suggest that h$\alpha$mr stars (high-$\alpha$, metal rich) may have originated from 
the inner disk (e.g. inner thick-disk members), they do not allow us to exclude the possibility that they 
represent a whole new Galactic population. More observations are needed to resolve this uncertainty.


\section{Planet-host stars in different stellar ''families``}

In our sample we have 135 planet hosts and 976 stars with no known orbiting planet. The left panel of Fig.~\ref{figure_2} shows the 
distribution of planet- host and non host stars in the [$\alpha$/Fe] vs [Fe/H] space. In the high metallicity region ([Fe/H] $>$ 0) the 
frequency of Jovian host stars which are also enhanced by $\alpha$ elements is higher (30.0 $\pm$ 8.3 \%) than the one for the hosts with low-$\alpha$
content (19.6 $\pm$ 2.2 \%).

\articlefiguretwo{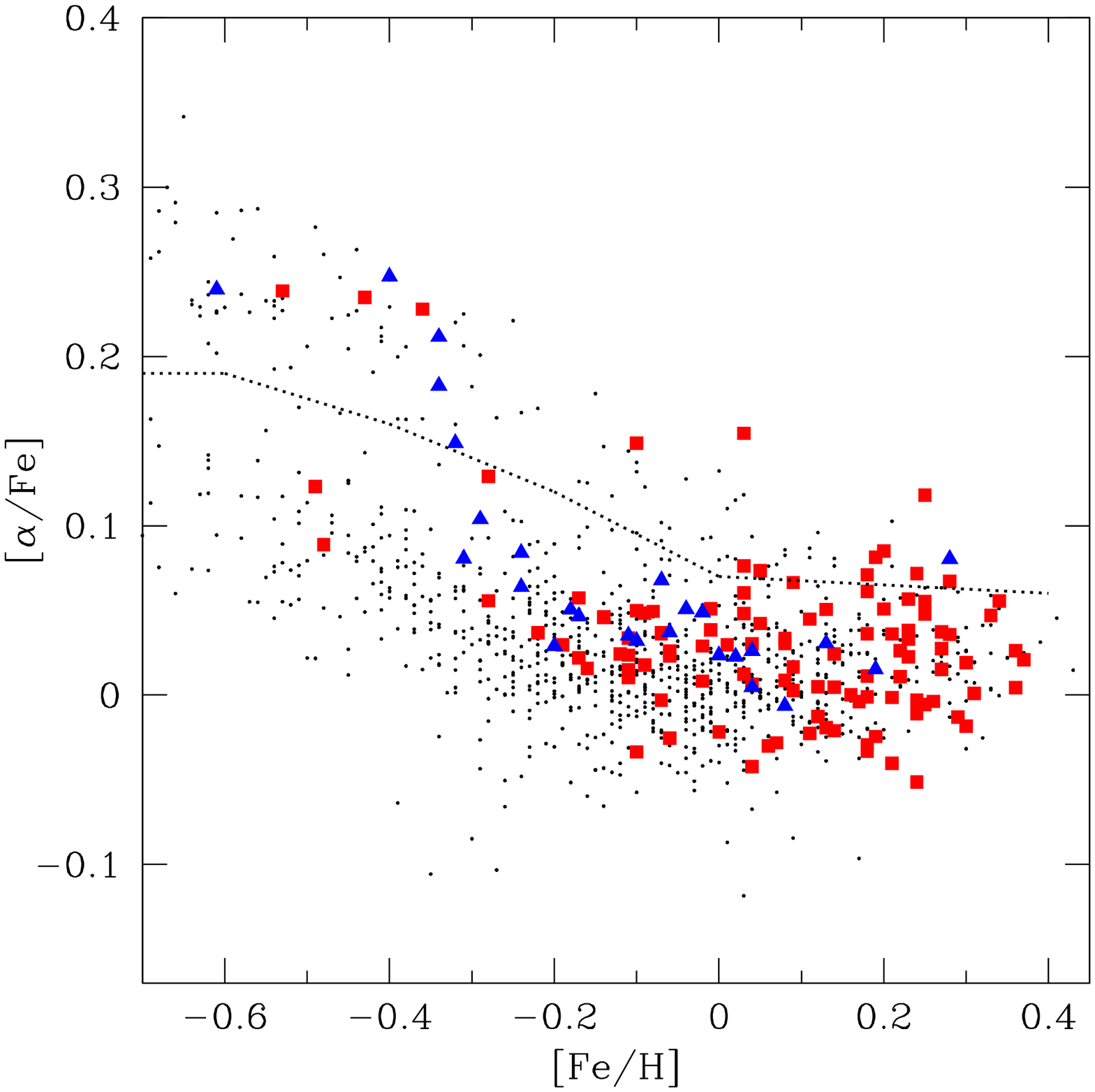}{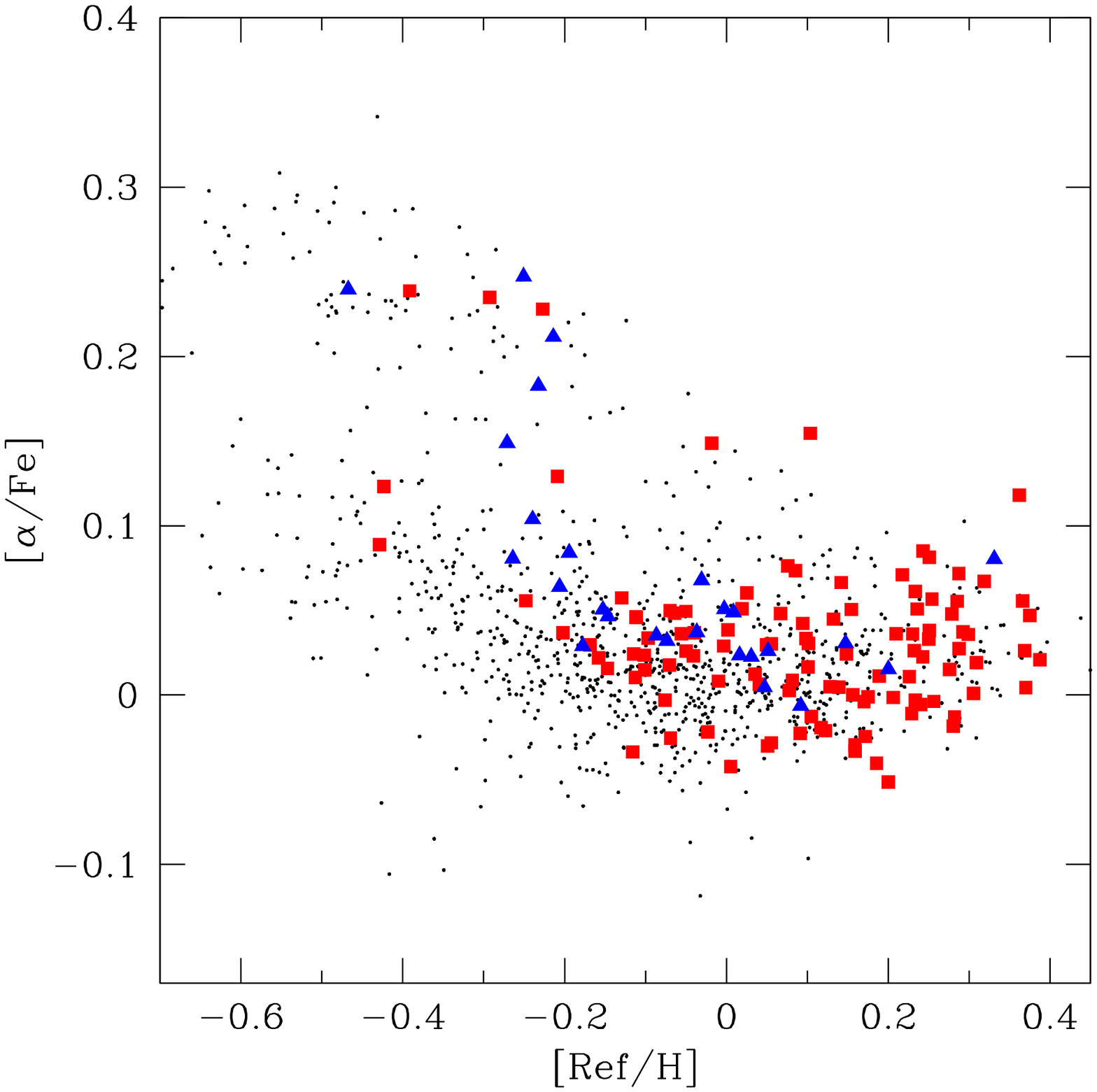}{figure_2}{[$\alpha$/Fe] versus [Fe/H] and [$\alpha$/Fe] versus [Ref/Fe] for the total sample. 
The symbols for non host stars are the same as in Fig.~\ref{figure_1}. Black dots represent the non-host stars, red squares refer to the Jovian hosts 
and blue triangles refer to the stars hosting exclusively Neptunians and super-Earths.}

Higher frequency rate of PHSs is observed in the ''very`` metal-poor region ([Fe/H] $\lesssim$ -0.3), 
where 8 hosts from 10 have high [$\alpha$/Fe] values (see also Adibekyan et al., 2012b submitted to A\&A).
The aforementioned $\alpha$-enhanced hosts are belonging to chemically defined ''thick`` disk, and the remaining 2 to the 
''thin`` disk. Interestingly, this thin/thick proportion changes dramatically when we apply a purely kinematic approach to separate the thin and thick discs.
The kinematical separation suggests that  6 or 3 host are from the thick disc, 4 or 6 stars belong to thin disc and 1 or 2 stars can be classified as 
transition stars depending on the kinematic criteria 
used - \citet{Bensby_03} or \citet{Reddy_06}, respectively.
Note that the plenty of Jpvian hosts in the thick disk were reported in \citet{Haywood_08,Haywood_09}. 

In the Fe-poor regime it is difficult to conclude what is the main reason that most of the PHSs lie in the high-$\alpha$/thick-disk region.
Following to \citet{Gonzalez_09}, and considering ''Ref`` index  which quantifies the mass abundances of Mg, Si and Fe,  instead of [Fe/H], 
we can see that the observed plenty of metal-poor PHSs in the thick disc ``disappears'': most metal-poor planet host stars in the thick disc 
have the same [Ref/H] distribution as their thin disc counterparts. Moreover, our data show that planet host stars start to have high [$\alpha$/Fe] ratios 
at lower metallicites when they still belong to the thin disc (see also Adibekyan et al., 2012b submitted to A\&A).

Our results allow us to suppose that the certain chemical composition and not the Galactic birth place is the determining factor that most of the 
metal-poor PHSs lie in the high-$\alpha$/thick-disk region. These results also suggest that in general, metals other than iron may also have an important 
contribution to planet formation if the amount of iron is insufficient to easily form a planet. Five stars out of 8 $\alpha$-enhanced metal-poor 
PHSs are orbiting low mass planets which indicates the importance of the $\alpha$ elements in the formation of Neptune-like planets in the metal-poor domain.

\acknowledgements {This work was supported by the European Research Council/European Community under the FP7 through Starting Grant agreement number 239953. 
N.C.S. also acknowledges the support from Funda\c{c}\~ao para a Ci\^encia e a Tecnologia (FCT) through program Ci\^encia\,2007 funded by 
FCT/MCTES (Portugal) and POPH/FSE (EC), and in the form of grant reference PTDC/CTE-AST/098528/2008. V.Zh.A., S.G.S. and P.F. also acknowledge
the grants reference SFRH/BPD/70574/2010, SFRH/BPD/47611/2008 and PTDC/CTE-AST/098528/2008  from FCT (Portugal), respectively.}

\bibliography{vadibekyan.bib}

\end{document}